\documentclass[12pt]{article}

\usepackage{amssymb, amsmath}
\usepackage{booktabs}

\usepackage{graphicx}
\usepackage{bbm}
   
\usepackage{a4wide}
\usepackage{xcolor}
   
\usepackage{multirow}
\usepackage{color}

\newcommand{\pr}{\ensuremath{\mathbb{P}}}
\newcommand{\Ex}{\ensuremath{\mathbb{E}}}
\newcommand{\Var}{\ensuremath{\mathbb{V}{\rm ar}}}

\newtheorem{remark}{Remark}

\graphicspath{./figures}

\begin{document}

\title{Modeling COVID-19 hospital admissions and occupancy in the Netherlands}


\author{Ren\'e Bekker$^{a,b}$, Michiel uit het Broek$^{a,c}$, and Ger Koole$^{a,b}$\\
\small$\strut^a$LCPS - Landelijk Co\"ordinatiecentrum Patiënten Spreiding, the Netherlands\\
\small$\strut^b$Department of Mathematics, Vrije Universiteit Amsterdam, the Netherlands\\
\small$\strut^c$Department of Operations, University of Groningen, the Netherlands}

\maketitle

\begin{abstract}
We describe the models we built for hospital admissions and occupancy of COVID-19 patients in the Netherlands. These models were used to make short-term decisions about transfers of patients between regions and for long-term policy making. We motivate and describe the model we used for predicting admissions and how we use this to make predictions on occupancy.
\end{abstract}


\section{Introduction}

The coronavirus has an enormous impact on our health system and today's society as a whole. On March 11, 2020, the World Health Organization has officially characterized COVID-19 as a pandemic. By the end of January 2021, the number of people diagnosed worldwide with COVID-19 crossed the 100 million mark \cite{WHO}, which has put a tremendous strain on scarce hospital capacities. Specifically, the pandemic places a load on clinical bed capacity, and in particular on Intensive Care Units (ICU's), that is well beyond the currently available bed capacities \cite{Covid2020, Covid2020b}.
The catastrophic situation in Lombardy, Italy, mid-March 2020 has tragically shown the impact of the lack of health capacities \cite{Rosenbaum2020}, and the need to manage hospital bed capacities as good as possible.
In \cite{Phua2020}, the authors call upon ICU practitioners, hospital administrators, governments,
and policy makers to be prepared early for a substantial increase in critical care capacity. 
Their recommendations relate to, among others, ICU capacity and ICU staffing.
More specifically, they recommend to make plans for an increase in capacity as a result of a rapid increase in critically ill COVID-19 patients.

The aim of this paper is to present a prediction model to support plans related to adjusting clinical bed capacities. 
In the Netherlands, there is a national plan in place for upscaling the number of (ICU) beds for COVID-19 patients. 
Moreover, to balance the pressure on clinical and ICU beds over the Netherlands, patients may be relocated to different regions; the LCPS (Landelijk Co\"ordinatiecentrum Pati\"enten Spreiding) is the Dutch center responsible for these relocations. 
In both situations, regions and local hospitals need a couple of days to modify the number of available COVID-19 beds, and thus require occupancy predictions of a couple of days ahead.   

The prediction model that we implemented consists of two steps. First, we predict arrivals on the basis of historical data. 
For this, we employ a linear programming model that is inspired by smoothing splines that incorporates weekly seasonality and requires little data. The prediction has interpretations in terms of the day-to-day reproduction factor. 
These arrival predictions, together with information about the length-of-stay (LoS) distribution, is used in the second step to predict hospital occupancy. This second step uses methods stemming from queueing theory, specifically from discrete-time infinite server queues. In this paper we use publicly available data to make predictions at the national level. 
The same model was and is used to make predictions at the regional level at the LCPS.  


The amount of literature on short-term predictions (in the order of days) of bed occupancy levels is limited. 
Recently, \cite{Farcomeni2020} used an ensemble of two forecasting methods for a short-term forecast of occupied COVID-19 beds in Italy. Furthermore, \cite{Zhao2020} applied epidemic models for short-term ICU occupancy forecasts in Switzerland; \cite{Massonnaud2020} uses a similar approach for the situation in France. 
The authors of \cite{Nikolopoulos2021} focus on a collection of countries and provide predictive analytic tools for excess demand in the supply chain due to COVID-19.
We refer to \cite{Guan2020} for an overview of the transmission dynamics of the COVID-19 pandemic.
The studies above focus directly on the number of occupied beds. We think that queueing-based insight is essential to understand the relation between arrivals and occupancy, which the studies above are lacking.  
The study of \cite{Massey-covid19} provides such a queueing-theoretic foundation. They explicitly focus on bed demand due to COVID-19 and use queueing models to present scenarios for the occupancy based on different arrival patterns of patients, that are based on different measures taken. However, the paper does not involve short-term predictions.
For a more extensive exposition of these types of queues in health care, we refer the reader to \cite{Worthington2020}.

There are also some papers focusing on short-term occupancy forecasts that is not directly related to COVID-19.
In \cite{Joy2005}, the authors use a hybrid approach of neural networks and ARIMA models to predict hospital occupancy directly. 
The studies \cite{Broyles2010, Kortbeek2015} mainly focus on hourly seasonality, and \cite{Littig2007} uses a predictive occupancy database, in which the data of each patient is registered. The focus of \cite{Davis2020, Pagel2017} is on distinguishing patient groups; see also \cite{Davis2020} for additional references. We note that our approach differs from the methods used in the papers mentioned above.

The organization of the paper is as follows. The method for predicting the arrivals is discussed in Section~\ref{sec:adm}. The LoS distribution can be found in Section~\ref{sec:los}, which is used to in the model to predict the bed occupancy in Section~\ref{sec:occ}. The prediction results can be found in Section~\ref{sec:pred}, whereas Section~\ref{sec:concl} concludes with a brief discussion on delayed care.

\section{Admissions} \label{sec:adm}
\begin{figure}
\includegraphics[width=\linewidth]{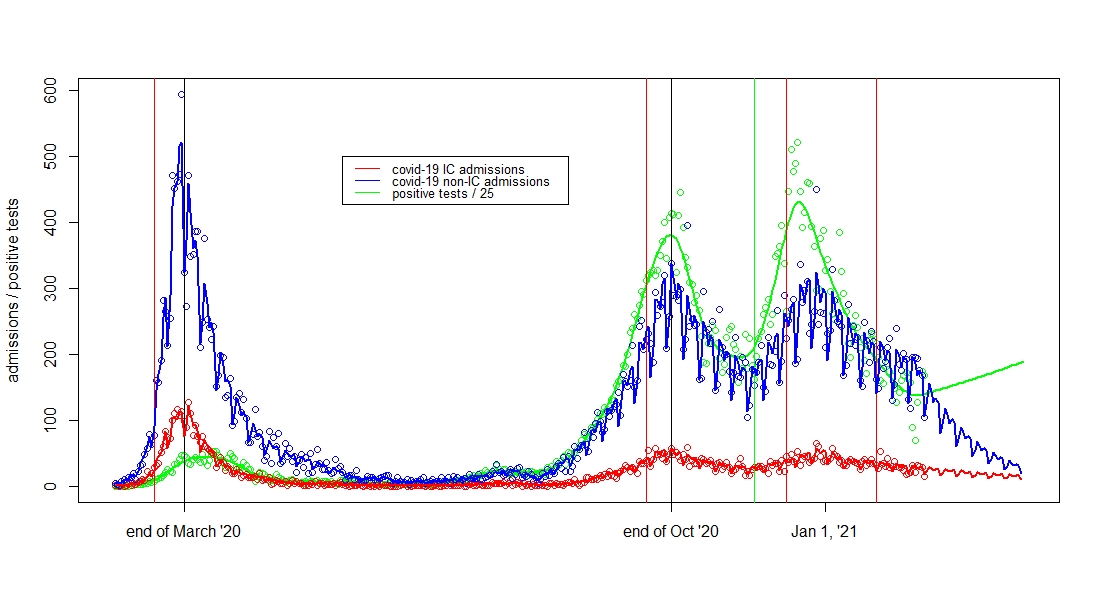}
\caption{Admissions and our fits and predictions; positive tests and their smoothing spline}
\label{spline-fig}
\end{figure}

There are different possibilities for building a model for admissions. An obvious option is a statistical model that uses historical values to make predictions. A disadvantage of such a model is that trend changes cannot be predicted. 
Also, many classical statistical models require a substantial number of observations in order to produce reliable predictions, whereas data is typically scarce when a new pandemic arises.
Furthermore, one would assume that somehow data on positive tests could be used, and that presumably positive tests occur before admissions, such that data on tests can be used to predict later admissions. 
In Figure~\ref{spline-fig} data on admissions and positive tests (at the day of registration) are plotted. 
Data comes from different publicly available sources, in this case NICE and RIVM, as is conveniently gathered at \cite{Zelst-github}.
The solid lines are smoothing splines on the logs of the values. 
The red bars indicate policy changes (partial lockdowns), it took 13 and 11 days for the numbers of admissions to go down (the black bars).
Surprisingly, the number of positive tests spikes at the same days as the number of admissions. For this reason, the number of newly registered positive tests per day cannot be used to predict trend changes in admissions. 
Another reason are the substantial changes in test policy and behavior. The green bar corresponds to one (of many) changes in test behavior; from that day (Dec 1) on civilians without symptoms could also get a test. 
We see that this led to a sharp increase in number of positive tests.
Hospital admission also increased, but at a lower pace.
Other variables than number of positive tests were tried as well on their ability to predict admissions, but they were neither useful.
For these reasons we focused on predicting admissions without external variables.

A statistical model for predicting daily admissions should have the following properties:
\begin{enumerate}
	\item it should be smooth but at the same time allow for trend changes;
	\item it should have non-negative predictions and exponential growth or 	decline;
	\item it should model the intra-week seasonality present in the data.
\end{enumerate}
For these reason we chose, inspired by smoothing splines, a model with the following features:
\begin{enumerate}
\item an additive model on the logs because of the multiplicative effect of time and the intra-week seasonality;
\item that minimizes a weighted sum of errors and second differences to have a smooth trend;
\item that uses absolute values to reduce the impact of outliers and few trend changes, hopefully representing the policy changes.
\end{enumerate}

A similar model is used in \cite{LeeuwenK-forecasting} to forecast demand in hospitality.
As the model is inspired by smoothing splines, it requires little data, which is preferable at the start of a pandemic. 
In mathematical terms, let~$a_t$ be the realization, either of the admissions at the ICU or the clinics. 
Our statistical models minimizes the sum of errors and trend changes, thus it is actually a minimization problem.
The decision variables are~$s_d$ and~$x_t$, the day factors and the weekly factors, respectively. 
Let $w(t)$ be the weekday of day~$t$, thus $s_{w(t)}$ is the day factor of day~$t$. 
Also define $\Delta x_t=x_t-x_{t-1}$ and $\Delta^2x_t=\Delta x_t - \Delta x_{t-1}$, and let $T$ be the last day with data.
Our minimization problem is:
\begin{equation} \label{eqn:lp-arr}
\min_{x,s}\sum_{t=1}^T|x_t+s_{w(t)}-\log a_t|+\lambda\sum_{t=3}^{T}|\Delta^2x_t|.
\end{equation}
Here, $\lambda\ge0$ is a parameter that determines the smoothness of the prediction, the ``smoothing parameter''.
The first term in (\ref{eqn:lp-arr}) gives the difference between the smoothed curve and the data and the second term introduces a penalty for trend changes.
The fit is given by $\exp (x_t+s_{w(t)})$, $t\le T$, whereas the $t$-day ahead prediction is
\begin{equation*}
\hat A_{T+t}  =  \exp \left( x_T+ t (x_T-x_{T-1})+s_{w(T+t)} \right). 
\end{equation*}

\begin{figure}
\includegraphics[width=\linewidth]{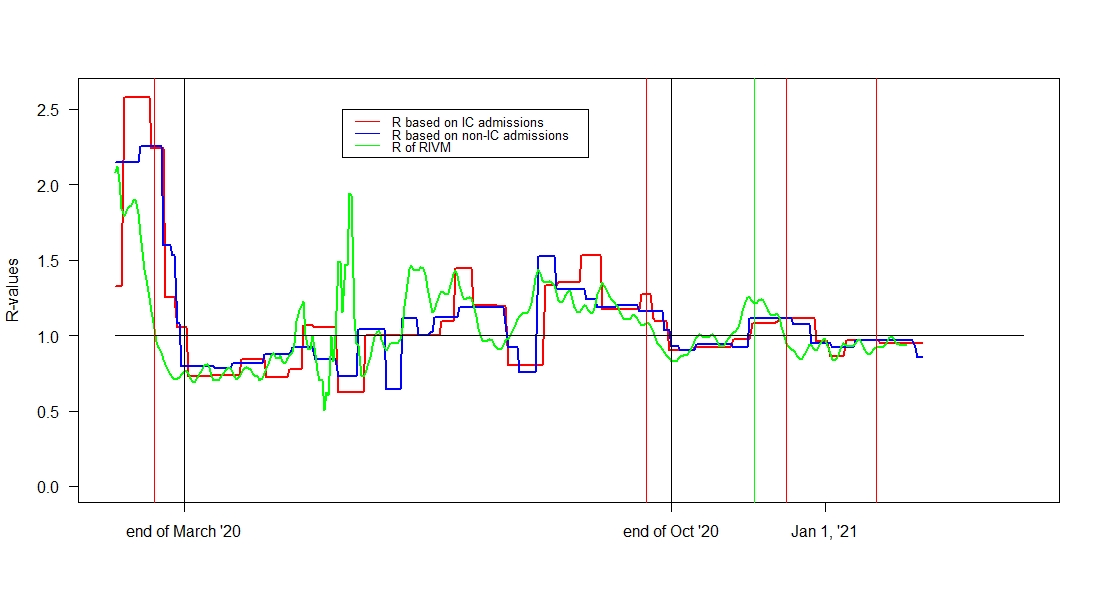}
\caption{Reproduction factors over time}
\label{r-fig}
\end{figure}

It is interesting to note that $r_t={\rm e}^{x_t-x_{t-1}}$ is the fractional de-seasonalized increase or decrease. 
It can be interpreted as a day-to-day ``reproduction factor". 
Epidemiologists define the reproduction factor~$R_t$ as the amount of people that get infected on average by one infected person at time~$t$.
As the incubation time is around four days there should be relation between~$R_t$ and~$r_t^4$. 
In Figure~\ref{r-fig}, $r_t^{4.5}$ is compared to~$R_t$ as it is determined by the Dutch National Institute for Public Health and the Environment (RIVM). 
We see a similar shape, and that the biggest correlation is for a lag of around twelve days, which corresponds roughly to the time between infection and hospital admission.

\section{Length of stay} \label{sec:los}

To determine the length of stay (LoS), we use data of NICE again. Specifically, on their website NICE presents data describing the frequencies of number of days that patients spend at the ICU and the clinic. 
Define $S$ as a random variable denoting the number of hospitalized days taking values in $\{0,1,\ldots\}$.
That is, $S$ may be interpreted as the number of overnight stays at the ICU or the clinic. Some recent studies \cite{Armony2015, Shi2016} have described the LoS at two time scales. The LoS in hours depends on many operational factors, whereas the LoS in days is attributed to medical factors. Our focus is on the latter, i.e., the time resolution in days. 

Currently, there are still COVID-19 patients present at the ICU and at the clinic, yielding right-censoring of the data. 
Clearly, the number of patients present is also non-negligible compared to the total number of COVID-19 patients, which in particular holds for the ICU. Therefore, to estimate the LoS distribution, we use the Kaplan-Meier estimator. 
In particular, we have $\hat \pr(S \ge 0) = 1$ and, for $t=1,2,\ldots$,  
\[
\hat \pr(S \ge t) = \prod_{s=1}^t \left( 1 - \frac{d_s}{n_s} \right),
\]	
where $d_s$ is the number of patients that are discharged after $s$ days, and $n_s$ is the number of patients that have a LoS of at least $s$ days (either discharged or still present). 

The mean and standard deviation of the LoS can be found in Table~\ref{tab:LoS}. 
We see that the average LoS at the ICU increases with over a day by taking the right-censoring into account. The impact is smaller at the clinic as a smaller fraction of the patients is still present (8.2\% at the ICU vs 3.4\% at the clinic).

\begin{table}
\centering
\begin{tabular}{l|ccc|ccc} \toprule
  	&	\multicolumn{3}{c|}{ICU}	& \multicolumn{3}{c}{Clinic}	 \\
  & \# patients & Mean & Stdev & \# patients & Mean & Stdev 	\\ \midrule
 Patients discharged or died	& 6,984 & 15.35 & 12.81 & 37,274 &  8.01 &  6.96 \\
 Patients currently treated		& 627 	& 17.25 & 13.62 & 1,323  & 11.98 & 12.92 \\
 Kaplan-Meier estimate			&  		& 16.64 & 13.69 & 		 &  8.43 &  7.51 \\ \bottomrule
\end{tabular}
\caption{LoS based on public NICE data}
\label{tab:LoS}
\end{table}

It is natural to consider the LoS at the time scale of minutes or hours, and model the LoS as a continuous random variable. 
There is also a considerable body of literature devoted to fitting probability distributions to such a continuous LoS.
Specifically, let $X$ represent a LoS taking values in $(0,\infty)$.
Recall that $S$ is a random variable denoting the number of hospitalized days taking values in $\{0,1,\ldots\}$.
When fitting a distribution to the LoS, we will use a fit to the continuous LoS $X$, and use a continuity correction to find the distribution of $S$. In particular, we have, for $t=1,2,\ldots$,
\[ \pr(S \ge t) = \pr(S > t-1) \approx \pr(X \ge t - 0.5). \]

In \cite{Armony2015}, a lognormal distribution is found to fit the LoS data well. The authors also pose the challenge to explain why lognormal distributions seem to fit service durations so well. Other common distributions for lengths of stay or survival functions are gamma and Weibull distributions \cite{Marazzi1998}; mixtures of exponentials may also be appropriate. We refer to \cite{Vekaria2020} for a study of the LoS of COVID-19 patients in the UK based on a Weibull distribution. 
In line with the LoS distribution of COVID-19 patients worldwide \cite{Rees2020}, we fit lognormal, gamma, and Weibull distributions. 
In Figures~\ref{fig:los-icu} and~\ref{fig:los-clinic}, these distributions are displayed together with the data adjusted by the Kaplan-Meier estimate.
For both the ICU and the clinic, the gamma and Weibull distributions can  hardly be distinguished. 
Interestingly, for the ICU the gamma and Weibull distributions provide visually excellent fits, whereas for the clinic the lognormal distribution provides very good fits.

\begin{remark}
There are different ways to determine parameters of our parametric distribution $X$. From the perspective of medical specialist and decision makers, the method of moments is especially appealing as the first two moment are relatively easy to interpret. For instance, the impact of changes in the LoS distribution are straightforward to incorporate. 
For $X \sim LogNormal(\mu, \sigma^2)$, we obtain 
$\mu = \ln\left( \bar x^2 / \sqrt{\bar x^2 + s^2} \right)$ and 
$\sigma^2 = \ln\left( 1+  s^2 / \bar x^2 \right)$,  with $\bar x$ and $s^2$ denoting the sample mean and the sample variance.
For $X \sim Gamma(\alpha, \beta)$, we obtain the shape parameter $\alpha = \bar x^2 / s^2$ and rate parameter $\beta = \bar x / s^2$. 
For Weibull distributions, there are no closed-form expressions when using the method of moments.
\end{remark}

\section{Occupancy} \label{sec:occ}

To predict the occupancy we use principles from queueing theory to describe the evolution of the number of COVID-19 patients. Essentially, we model the number of patients as a (discretized) infinite-server queueing model with a time-dependent arrival pattern. For the special case of (continuous) time-dependent Poisson arrivals, the $M_t/G/\infty$ has well been analyzed with tractable results \cite{Bekker2010, Eick1993, Feldman2008}; \cite{Massey-covid19} uses such an $M_t/G/\infty$ model to quantify how flattening the curve affects peak demand for hospital beds. 
The application of infinite-server models, also in discrete time, is also discussed in \cite{Worthington2020}. 
As our goal is to predict the demand for beds without capacity constraints, the infinite-server assumption is appropriate, albeit we use a discrete-time version.

When predicting future occupancy, we need to distinguish two groups of patients: (i) patients that are currently present, and (ii) patients that will arrive in the future. For the patients that will arrive in the future, we need a prediction of admissions (as described in Section~\ref{sec:adm}) and the subsequent length of stay (as described in Section~\ref{sec:los}). For the first group, observe that the patients that are currently present, the total length of stay differs from the one in Section~\ref{sec:los} whereas part of the length of stay has elapsed. Since we predict on publicly available data, we cannot use the elapsed length of stay of each individual patient. A reasonable alternative seems to use the stationary residual length of stay (for which $\pr(S^r \ge t) = \sum_{k=t}^{\infty} \pr(S > k)/\Ex S$), which follows directly from renewal theory. A disadvantage of the stationary residual length of stay is that the arrival process is obviously not stationary. Therefore, we propose an alternative that takes the past arrival pattern into account.

Next, we derive the residual length of stay $S^r$ of a tagged patient present at time $T$. Note that the probability that this patient arrived at day $T-u$ is proportional to $a_{T-u} \pr(S \ge u)$, for $u=1,\ldots,T$. Hence, the probability that this tagged patient arrived at day $T-u$ is
\[ \frac{a_{T-u} \pr(S \ge u)}{\sum_{k=1}^T a_{T-k} \pr(S \ge k)}. \]
The probability that the residual length of stay of the tagged patient is at least $s$, when the patient arrived at day $T-u$, equals $\pr(S \ge s+u \mid S \ge u) = \pr(S \ge s+u) / \pr(S \ge u)$. Combining the above, we have
\begin{equation*}
\pr(S^r \ge t) = \sum_{u=1}^T  \frac{a_{T-u} \pr(S \ge u)}{\sum_{k=1}^T a_{T-k}  \pr(S \ge k)} \frac{ \pr(S \ge t+u)}{ \pr(S \ge u)}
 = \frac{\sum_{u=1}^T a_{T-u} \pr(S \ge t+u)}{\sum_{k=1}^T a_{T-k}  \pr(S \ge k)}.
\end{equation*}
Observe that this is consistent with the stationary residual length of stay by taking $a_{\cdot}$ constant and letting $T \to \infty$.

Now, we turn to predicting the occupancy. 
As the allocation of COVID-19 patients is based on the occupancy in the morning, we focus on $N_t$, the number of occupied beds at the beginning of day $t$. We then have the following relation
\[ N_{T+t} = \sum_{i=0}^{N_T} \mathbbm{1} \{ S_i^r \ge t \} + \sum_{s=0}^{t-1} \sum_{i=1}^{A_{T+s}} \mathbbm{1} \{ S_i \ge t-s\}, \]
where $S_i^r$ is the residual LoS of the $i$th patient present, and $S_i$ represents the LoS of the $i$th patient arriving on that specific day.
The first term is due to patients that are currently present at time~$T$, whereas the second term are patients arriving in the future.
Observe that with the relation above it is possible to derive the distribution of $N_{T+t}$. Focusing on the expectation, it holds that
\begin{equation*}
\hat N_{T+t} = \Ex N_{T+t} = N_T \pr(S^r \ge t) + \sum_{s=0}^{t-1} \Ex A_{T+s} \pr(S \ge t-s),
\end{equation*}
providing the $t$-day ahead prediction $\hat N_{T+t}$ at day $T$.
Moreover, using the same relation above and assuming that $A_{T+s}$ and $A_{T+u}$ are independent for $s \neq u$, the variance is 
\begin{eqnarray*}
 \Var N_{T+t} &=& N_T  \pr(S^r \ge t) (1- \pr(S^r \ge t)) \\
 & & \;\; + \sum_{s=0}^{t-1} \left( \Var A_{T+s} \times \bar G(t-s)^2 + \Ex A_{T+s} \times \bar G(t-s)(1-\bar G(t-s)) \right),
\end{eqnarray*}
where $\bar G(t-s) = \pr(S \ge t-s)$. Note that the expression above simplifies if the arrivals follow a Poisson process with a known parameter. In that case $\Var A_{T+s} = \Ex A_{T+s}$ and $\Var N_{T+t}$ will converge to $\Ex N_{T+t}$, such that $N_{T+t}$ will behave as a Poisson random variable for $t$ large enough.

\section{Predictions} \label{sec:pred}

In this section we present the numerical results that follow from our prediction model. As we only have reliable occupancy data of COVID-19 patients from mid October 2020, we will use the time period from November 1, 2020, until February 1, 2021 as an illustration. This also involves an interesting period due to the remarkable behavior of infections and hospital admissions during the `second wave'. 
In line with the operations at the LCPS, we use predictions of 3 and 7 days ahead. 
To assess the accuracy of the predictions, we use the following three evaluation measures: weighted absolute percentage error (WAPE), mean absolute error (MAE), and root mean squared error (RMSE). We note that the WAPE is also referred to as the weighted MAPE. For a period of $n$ days, these measures are defined as
\begin{equation*}
{\rm WAPE} = \frac{\sum_{t=1}^n \left| y_t - \hat y_t \right|}{\sum_{t=1}^n y_t}, \quad
{\rm MAE} =  \frac{1}{n} \sum_{t=1}^n \left| y_t - \hat y_t \right|, \quad
{\rm RMSE} = \frac{1}{n} \sqrt{\sum_{t=1}^n (y_t - \hat y_t)^2},
\end{equation*}
where $y_t$ and $\hat y_t$ are the actual and predicted values, respectively, at day $t$.

\begin{figure*}
	\centering
	\includegraphics[width=0.9\textwidth]{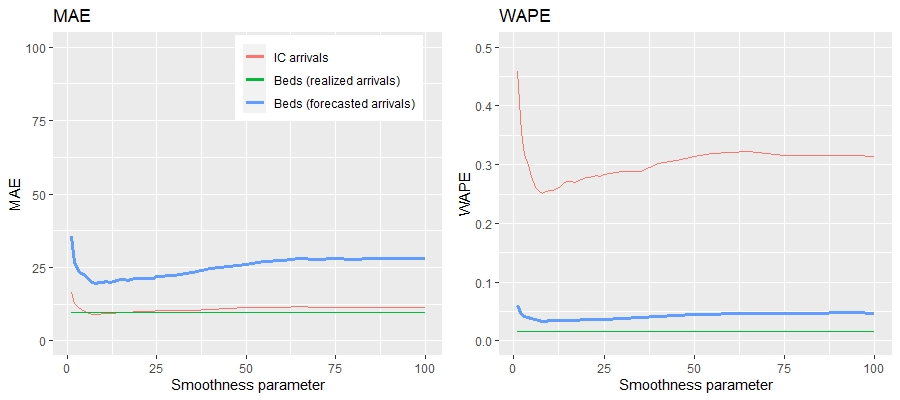}
	\caption{Impact of the smoothing parameter $\lambda$ on accuracy of arrival and occupancy 3-day ahead predictions at the ICU} 
	\label{fig:smoothing}
\end{figure*}

First, in the arrival predictions (\ref{eqn:lp-arr}) there is a tunable smoothing parameter $\lambda$. Figure~\ref{fig:smoothing} shows the impact of the smoothing parameter on the WAPE and MAE for both the ICU arrivals (red lines) and occupied ICU beds 3-day ahead predictions. For the ICU beds, we use both the complete prediction model (blue lines), and the occupancy model fed by the actual arrival stream (green lines). The aim of the latter is to obtain insight in the impact of the LoS on the accuracy of the occupancy forecast, as there is no error in the arrival prediction in that case. 
This also implies that the green lines are not affected by $\lambda$ as this parameter only affects the arrival prediction.

The differences between the green and blue lines should be interpreted as the error in occupancy prediction that is due to the unknown arrival process. Also observe that the arrivals (red) and occupancy (blue) are at a completely different level, as will also become apparent below, explaining the differences in absolute (MAE) and relative (WAPE) errors. 	
Clearly, for $\lambda$ very small the forecast is too responsive, whereas the opposite occurs for large $\lambda$. We note that the behavior is similar for 7-day ahead predictions and for the clinic.
In practice, it is desirable to tune the parameter $\lambda$ based on contextual information, such as measures taken, as this may improve the prediction \cite{Sanders2001}.
For consistency, we use a single smoothing parameter of $\lambda = 10$ in the experiments below.

\begin{figure*}
	\centering
	\includegraphics[width=0.95\textwidth]{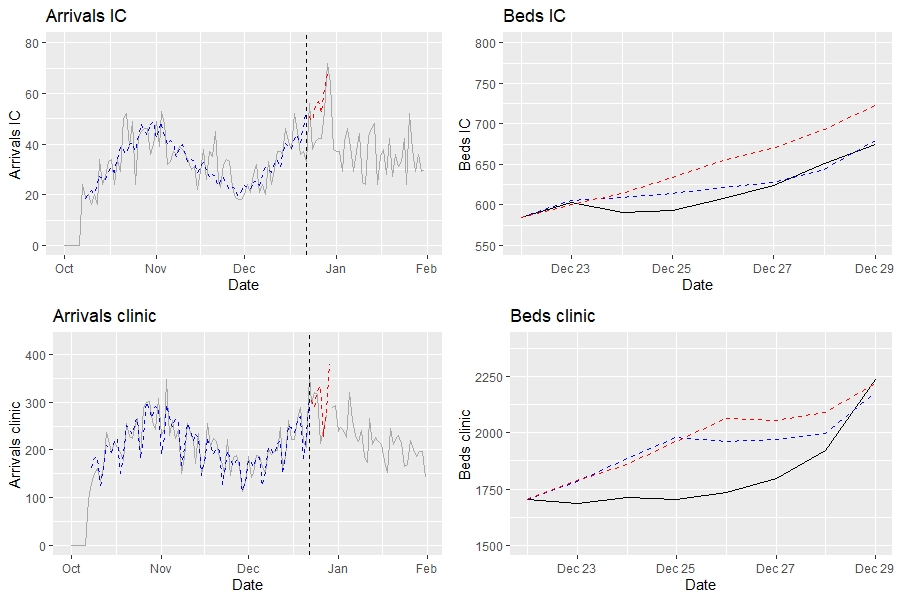}
	\caption{Predictions of arrivals (left) and occupancies (right) for the ICU (top) and clinic (bottom) at December 22, 2020} 
	\label{fig:example1}
\end{figure*}

Next, we visualize the predictions for $1,\ldots,7$ days ahead for the arrivals and occupancy of both the ICU and the clinic. In Figure~\ref{fig:example1} we present the predictions made at December 22, 2020. The arrivals are plotted on the left, with the solid lines the actual values, the blue dotted lines the fit, and the red dotted lines the predictions. The occupancies are plotted on the right, with the solid lines the actual values again, the red dotted lines the predicted values, and the blue dotted lines the predictions when the arrivals are known; the aim of the latter is to obtain insight in the impact of inaccurate predictions for the arrival process.  

For the arrivals, we see a very good fit (blue line), with an apparent weekly arrival pattern, in particular for the clinic. The arrival predictions for the clinic are accurate, but for the ICU the model seems to overestimate the number of arrivals. 
Specifically, the increasing trend does not continue as strongly as suggested by the data up to Dec 22.
This also leads to an overestimation of the number of occupied ICU beds (compare the red line with the blue line for the ICU beds). Regarding the occupancy for the clinic, there seems an overestimation of the number of occupied beds for the period from Dec 24 until Dec 28. This is not due to the arrival predictions, as the red and blue lines are rather similar. It seems likely that some patients might be discharged earlier from the clinic in the period around Christmas. 


\begin{figure*}
	\centering
	\includegraphics[width=0.95\textwidth]{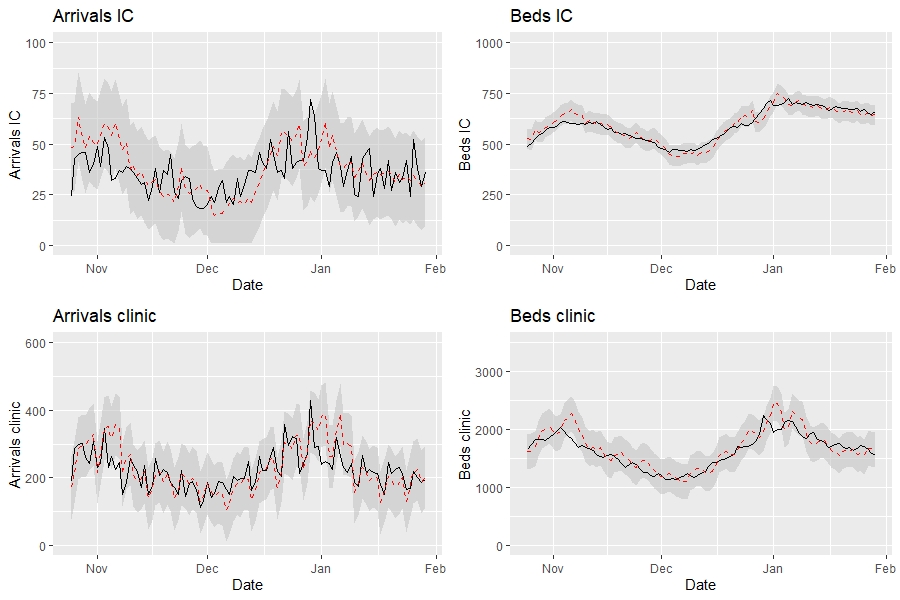}
	\caption{3-day ahead predictions of arrivals (left) and occupancies (right) for the ICU (top) and clinic (bottom)} 
	\label{fig:3day}
\end{figure*}
\begin{figure*}
	\centering
	\includegraphics[width=0.95\textwidth]{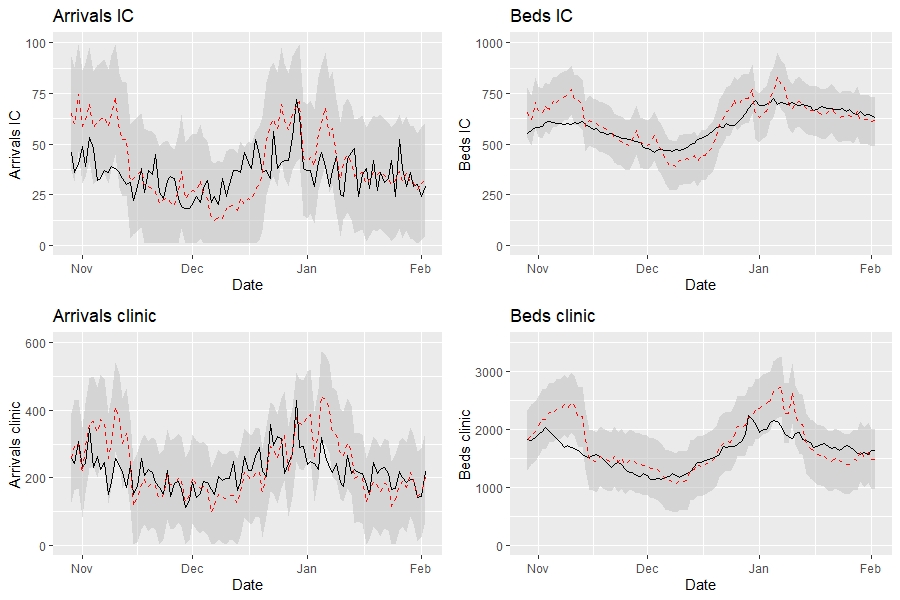}
	\caption{7-day ahead predictions of arrivals (left) and occupancies (right) for the ICU (top) and clinic (bottom)} 
	\label{fig:7day}
\end{figure*}

To see how the predictions behave over time, we use a rolling horizon and, for every day, make predictions for 3 and 7 days ahead.
In Figure~\ref{fig:3day} the 3-day ahead predictions (with corresponding bandwidth) together with their realizations are shown for the arrivals and occupancies for the ICU and the clinic. Overall, the predictions are visually accurate. We see that the predictions tend to deviate from the realizations at moments when the arrival pattern changes, i.e., when the arrivals reach a local peak or valley. When the number of arrivals is at such a local peak or valley, it takes a couple of days for the arrival prediction to detect that the local trend is changing, and this change is not caused by some random realizations. When the predictions are completely based on the time series (without further contextual information), it seems difficult to overcome such an issue. However, the prediction model is able to adapt to such trend changes after a couple of days.  

Similar 7-day ahead predictions are shown in Figure~\ref{fig:7day}. We see similar phenomena for the 7-days ahead predictions as for the 3-days ahead predictions. However, the bandwidth is wider and it naturally takes more time to detect a trend change for the 7-days ahead predictions than for 3-days ahead.

\begin{table}
\centering
\begin{tabular}{l|ccc|ccc} \toprule
  	&	\multicolumn{3}{c|}{3 days ahead}	& \multicolumn{3}{c}{7 days ahead}	 \\
  & WAPE & MAE & RMSE & WAPE & MAE & RMSE 	\\ \midrule
 Arrivals IC 						& 26\% &  9.19	&  11.27 & 34\%	&  12.10	&  14.90 \\
 Arrivals clinic 					& 15\% & 34.24 &  47.28 & 24\% &  52.02	&  68.55 \\
 Beds IC (realized arrivals) 		&  2\% &  9.72 &  12.67 &  2\% &  13.24	&  17.18 \\
 Beds IC (forecasted arrivals) 		&  3\% & 20.02 &  25.58 &  9\% &  51.02	&  63.64 \\
 Beds clinic (realized arrivals) 	&  6\% & 90.63 & 107.97 &  7\% & 106.97	& 126.73 \\
 Beds clinic (forecasted arrivals) 	&  8\% &126.25 & 162.20 & 13\%	& 216.90	& 290.65 \\ \bottomrule
\end{tabular}
\caption{Accuracy measures of arrival and occupancy predictions}
\label{tab:accuracy}
\end{table}

In Table~\ref{tab:accuracy} the accuracy measures of the predictions are presented, again for the arrivals and occupancies, and the ICU and the clinic.
Clearly, the relative errors (WAPE) are largest for the admissions, which is partly explained by the fact that the number of arrivals is considerably smaller than the number of occupied beds; see also Remark~\ref{rm:poisson_error} for the impact of scale. Moreover, it reveals that predicting arrivals is complicated for such a volatile process including changes in trend. The 3-day ahead prediction in the required number of ICU beds is remarkably accurate. Given the inherent randomness in the bed census process, see Remark~\ref{rm:poisson_error}, a WAPE of 3\% seems to be the best achievable. For the 7-day ahead prediction of ICU occupancy, we see that the error is mainly determined by the error in the arrival process (9\% with forecasted arrivals vs 2\% with actual arrivals).
Overall, the model performs very well for the most important predictions, i.e., the ICU occupancies.
Compared to the ICU, the predictions for the clinic occupancies seem not as good as expected. In particular, even with the actual arrival streams, the WAPE is still 6\% and 7\% for 3 and 7 days ahead, respectively. These errors can be explained by the discharge behavior at the clinic, where there are only few discharges during the weekend (which are compensated during the week).
We like to emphasize that the discharge behavior during the week only has a modest impact on the prediction results in our current practice, as the predictions are only used for at specific days during the week.


\begin{figure*}
	\centering
	\includegraphics[width=0.95\textwidth]{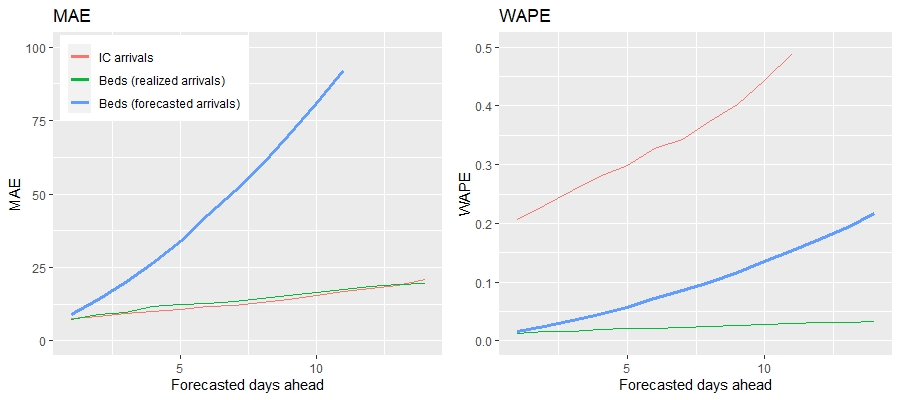}
	\caption{Accuracy of ICU predictions for $1,\ldots,14$ days ahead} 
	\label{fig:days_ahead}
\end{figure*}

Finally, we consider the impact of the number of days ahead on the accuracy (MAE and WAPE) of the ICU predictions in Figure~\ref{fig:days_ahead}. The red line concerns the arrivals, whereas the blue line is the prediction of the occupancy; the green line is the occupancy in case the actual arrivals are used (and deviations are due to the LoS). As the scale differs between arrivals and occupancy, the MAE is considerably smaller and the WAPE considerably larger for the arrivals compared to the occupancy. Of course, the predictions become less accurate when the forecast is longer ahead. If the actual number of arrivals are known, we see that the occupancy predictions (green line) remain quite accurate even for 14 days ahead. Hence, prediction of the arrival process is crucial, in particular for predictions that are more than a week ahead.

\begin{remark} \label{rm:poisson_error}
The assessment of the accuracy of predictions is complicated by the inherent randomness in arrivals and LoS. For instance, suppose that our aim is to predict the value of a Poisson random variable with rate $\mu$; the Poisson distribution typically reflects the randomness in arrivals or occupancy. The most accurate prediction would be $\hat y_t = \mu$. In that case, with $n \to \infty$ and using \cite{Crow1958}, we have ${\rm MAE} = 2 \mu^{\lfloor \mu \rfloor+1} e^{-\mu}/\lfloor \mu \rfloor !$, ${\rm WAPE} = {\rm MAE}/\mu$, and ${\rm RMSE} = \sqrt{\mu}$. For example, for $\mu$ equal to 50, 500, and 2000,
the MAE is 5.6, 17.8, and 35.7, respectively, 
whereas the WAPE is 11.3\%, 3.6\%, and 1.8\%, respectively. 
\end{remark}

\section{Conclusion and discussion} \label{sec:concl}

In this paper, we presented a mathematical model to give short-term predictions, in the order of days, of the number of occupied ICU and clinical beds due to COVID-19. The model first predicts the arrivals and then employs a queueing-based method to convert arrivals into occupancy. The predictions for the ICU occupancies are accurate, in particular for 3 days ahead. For the clinical occupancies, there is a seasonal component in discharges, with considerably less discharges during the weekend, that affects the performance of the predictions averaged over all days. An interesting topic for further research is to take the seasonal component in discharges into account as well.

Predictions of a couple of days ahead are crucial to properly manage ICU and clinical bed capacity and to relocate patients across the country. In essence, the framework is also suitable for longer-term scenarios, but an appropriate approximation of the behavior of the arrival process is then crucial. 
Moreover, COVID-19 admissions consume a considerable part of the resources at the ICs and clinics in the Netherlands. Additional resources were also used, such as post-anesthesia recovery beds, and anesthesiologists who worked as buddies next to the intensivists. This also reduced other forms of hospital capacity, together leading to reduced capacity for other forms of care leading to waiting lists for multiple forms of care. It is hard to quantify the impact of the delays. For example, \cite{RIVM-delayedcare} reports up to 50000 ``healthy years of life lost" due to the first wave, based on 28\% of the specialist medical care. However, some of this loss can be recovered if extra treatments are provided in the future. There is no centralized information on the length of waiting lists and the rate at which lives are lost.

From a mathematical view, it is interesting to study the impact of the second wave on the delayed care. For the moment the daily admissions have not reached the peak level of the first wave, but the rise and decline of the second wave has been much slower, leading to a higher number of patients and days of hospitalization. This inevitably leads to more delayed care, it is highly likely that waiting lists will become at least twice as long. This has a quadratic impact on the years of life lost: if twice as many patients wait on average twice as long before treatment, the total impact is 4 times higher. 
This amplifies the need for an efficient use of resources and good predictions of required capacity.

\paragraph{Acknowledgments}

Part of the work has been carried out during the period that we were affiliated with the LCPS. We would like to thank Marcel de Jong and the LCPS for providing us insight in the management of COVID-19 in the Netherlands and the pleasant and fruitful cooperation.

\appendix

\section{Length of stay distributions}

\begin{figure*}
	\centering
	\includegraphics[width=0.8\textwidth]{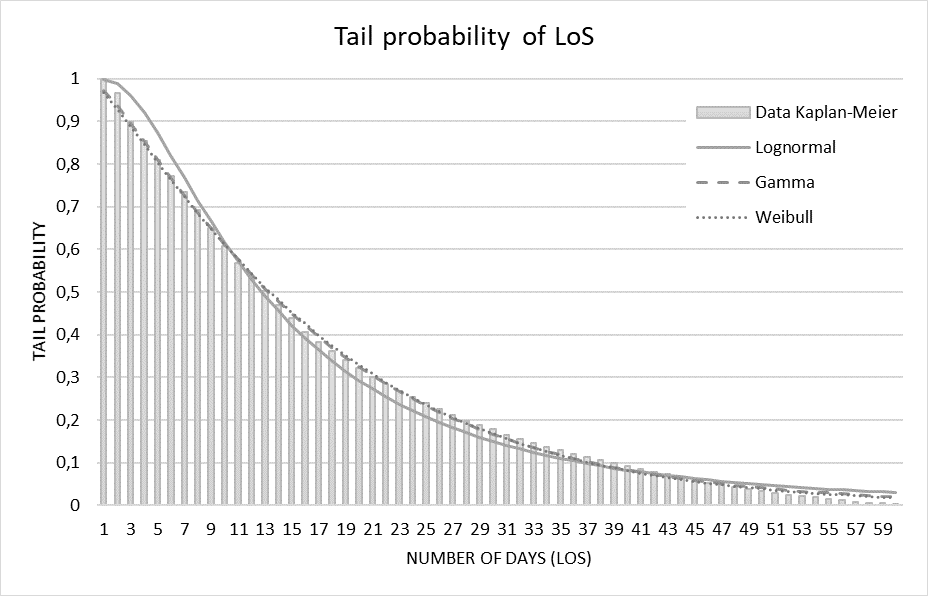}
	\caption{Tail distribution of the LoS at the ICU} 
	\label{fig:los-icu}
\end{figure*}

\begin{figure*}
	\centering
	\includegraphics[width=0.8\textwidth]{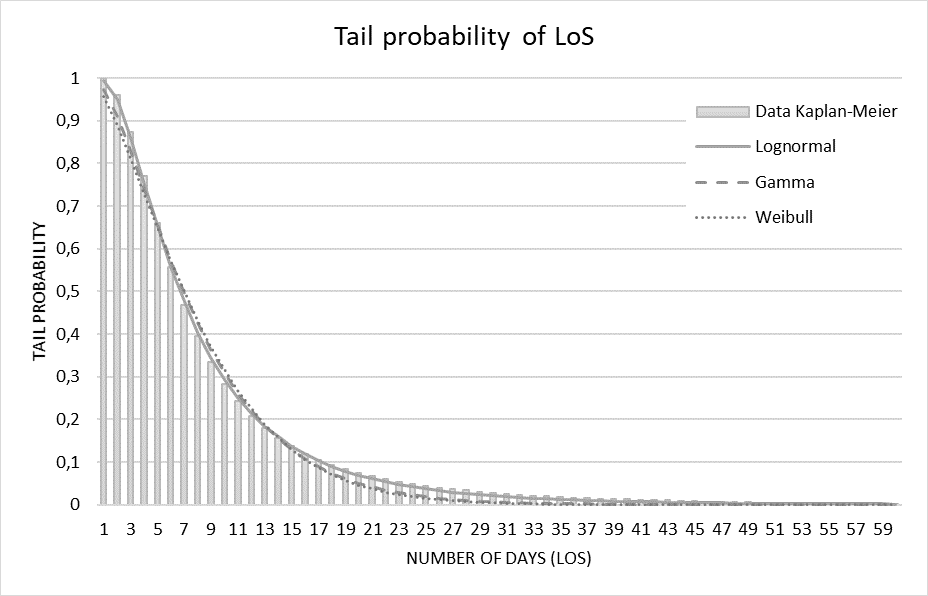}
	\caption{Tail distribution of the LoS at the clinic} 
	\label{fig:los-clinic}
\end{figure*}

In Figures~\ref{fig:los-icu} and~\ref{fig:los-clinic}, the length of stay distribution is displayed for the ICU and the clinic, respectively. For both cases, the data is plotted after applying the Kaplan-Meier estimator, together with lognormal, gamma, and Weibull fits.


\begin{small}

\end{small}
\bibliographystyle{plain}

\end{document}